\documentclass[12pt,a4paper]{article} 
\usepackage[latin2] {inputenc}
\usepackage{amsmath} 
\usepackage{amsfonts} 
\usepackage{amssymb} 
\newcommand{\idxrangeb}[3]{\ensuremath{{#1}_{#2}\ldots{}{#1}_{#3}}}
\newcommand{\idxrange}[2]{\ensuremath{\idxrangeb{#1}{1}{#2}}}
\newcommand{\labelb}[1]{\label{#1}}
\newcommand{\idxspc}{}
\newcommand{\refb}[1]{(\ref{#1})}
\newcommand{\bm}[1]{\mbox{\boldmath$#1$}}

\begin{document}

\author{M. Cvitan\thanks{E-mail: mcvitan@phy.hr},
  S. Pallua\thanks{E-mail: pallua@phy.hr} and
  P. Prester\thanks{E-mail: pprester@phy.hr}\\[5mm]
  \normalsize \it Department of Theoretical Physics,\\[-1mm]
  \normalsize \it Faculty of Natural Sciences and Mathematics,\\[-1mm]
  \normalsize \it University of Zagreb,\\[-1mm]
  \normalsize \it Bijeni\v{c}ka c. 32, pp. 331, 10000 Zagreb, Croatia}

\date{\normalsize Preprint ZTF 02-04}
\title{Entropy of Killing horizons from Virasoro algebra in $D$-dimensional extended Gauss--Bonnet gravity}

\maketitle

\begin{abstract}
We treat $D$-dimensional black holes with Killing horizon for extended Gauss--Bonnet gravity. We use Carlip method and impose boundary conditions on horizon what enables us to identify Virasoro algebra and evaluate its central charge and Hamiltonian eigenvalue. The Cardy formula allows then to calculate the number of states and thus provides for microscopic interpretation of entropy.
\end{abstract}

\section{Introduction}
The purpose of this paper is to investigate how some recent results on microscopic interpretation of black hole entropy depend on the form of gravity action.
The problem of microscopic description of black hole entropy was approached by different methods, like e.g. string theory, which treated extremal and near extremal black holes 
\cite{Strominger:1996sh}, 
or e.g., loop quantum gravity \cite{Ashtekar:1997yu}. 
Another line of approach to this problem is based on conformal field theory and Virasoro algebra. 
Such an algebra was identified by Brown and Henneaux \cite{BroHen86} in $2+1$ dimensions and after requiring asymptotic $AdS_3$ symmetry. The well known Bekenstein--Hawking entropy formula for Einstein gravity black holes was then reproduced \cite{BanStro}. In fact, there are essentially two independent approaches based on conformal field theory.
One particular formulation was due to Solodukhin who reduced the problem of $D$-dimensional black holes to effective two dimensional theory with fixed boundary conditions on horizon. This effective theory admits Virasoro algebra near horizon and calculation of its central charge allows to compute the entropy
\cite{Solod99,Carlip02}.
Another approach based on conformal field theory was developed by Carlip \cite{Carlip02,Carlip01,Carlip98,Carlip99}.
In fact Carlip has shown that under certain simple assumptions on boundary conditions near black hole horizon, one can identify a subalgebra of algebra of diffeomorphisms, which turns out to be Virasoro algebra.
The fixed boundary conditions give rise to central extension of this algebra. The entropy is then calculated from Cardy formula \cite{Cardy}
\begin{equation}\labelb{eqI1}
S_{\mathrm{C}}=2\pi\sqrt{\left( \frac{c}{6} - 4\Delta_{g}\right)\left( \Delta - \frac{c}{24} \right)  }\thickspace\textrm{,}
\end{equation} 
where $\Delta$ is the eigenvalue of Virasoro generator $L_0$ for the state we calculate the entropy and $\Delta_{g}$ is the smallest eigenvalue.
The corresponding entropy reproduces Bekenstein--Hawking formula \cite{BekHaw}
\begin{equation}
S_{\mathrm{BH}}=\frac{A}{4\pi G}
\thickspace\textrm{,}
\end{equation} 
where $A$ is area of the black hole horizon.
Several aspects of diffeomorphism algebra at horizon and its applications have been discussed in various papers \cite{Silva02}. 
Also interesting connected developments have been proposed in algebraic approach \cite{Gour:2002pj}. 
Till now, similar analysis was done for Einstein gravity and for dilaton gravity \cite{Carlip99,Jing:2000yn,Carlip02}.
Here, we shall consider Gauss--Bonnet generalization of Einstein gravity in $D$-dimensions \cite{Lovelock}. 
In fact it is known \cite{genent,Iyer:1994ys} that classical entropy differs from area law in Einstein theories and that for more general diffeomorphism invariant theory the entropy of black hole with bifurcate horizon is
\begin{equation}\labelb{eqS8E} 
S=-2\pi\int_{\mathcal{H}} {\bm{\hat\epsilon}\, E_{R}^{abcd}\eta_{ab}\eta_{cd}}
\thickspace\textrm{.}
\end{equation} 
Here, $\mathcal{H}$ is a cross section of the horizon, $\eta_{ab}$ denotes binormal to $\mathcal{H}$ and $\bm{\hat\epsilon}$ is induced volume element on $\mathcal{H}$.
The tensor $ E_{R}^{abcd} $ is given with
\begin{equation}\labelb{eqdefE}
E_{R}^{abcd}=\frac{\partial L}{\partial{R_{abcd}}}
\thickspace\textrm{.}
\end{equation} 
The tensor $ E_{R}^{abcd} $ has all symmetries of Riemann tensor $ R^{abcd} $. 
In this paper we shall treat Gauss--Bonnet gravity with Lagrangian density
\begin{equation}\labelb{igb}
L=
-\sum_{m=0}^{[D/2]}\lambda_m L_m(g)
\thickspace\textrm{.}
\end{equation}
Here, $[D]$ denotes integer part of $D$. The $m$-th density is
\begin{equation}\labelb{lgbm}
L_m(g)=\frac{(-1)^m}{2^{m}}
\delta_{a_{1}b_{1}\ldots{}a_{m}b_{m}}
      ^{c_{1}d_{1}\ldots{}c_{m}d_{m}}
{R^{a_{1}b_{1}}}_{c_{1}d_{1}}\cdots
{R^{a_{m}b_{m}}}_{c_{m}d_{m}}
\thickspace\textrm{,}
\end{equation}
where $\delta_{a_1\ldots{}a_k}^{b_1\ldots{}b_k}$ is totally antisymmetric product of $k$ Kronecker deltas,
normalized to take values 0 and $\pm 1$.
First few densities are
$L_0=1$, $L_1=-R$, $L_2=R_{abcd}R^{abcd}-4R_{ab}R^{ab}+R^2$.
Corresponding tensor $ E_{R}^{abcd} $ reads
\begin{equation}\labelb{eqdefEGB}
{E_R^{cd}}_{ab}=-\sum_{m=0}^{[D/2]}m\lambda_m\frac{(-1)^m}{2^{m}}
\delta_{a_{\idxspc}b_{\idxspc}a_{2}b_{2}\ldots{}a_{m}b_{m}}
      ^{c_{\idxspc}d_{\idxspc}c_{2}d_{2}\ldots{}c_{m}d_{m}}
{R^{a_{2}b_{2}}}_{c_{2}d_{2}}\cdots
{R^{a_{m}b_{m}}}_{c_{m}d_{m}}
\thickspace\textrm{.}
\end{equation} 
In particular, for Lagrangian which includes densities up to second order, 
tensor $ E_{R}^{abcd} $ reads
\begin{eqnarray}
E^{abcd} &=&
\left( \frac{1}{16\pi} + 2\alpha R \right) 
\frac{1}{2} 
\left( g^{ac}g^{bd}-g^{ad}g^{bc} \right) \nonumber\\
&+&2\alpha\left( R^{abcd} - g^{bd}R^{ac} + g^{ad}R^{bc} - g^{ac}R^{bd} + g^{bc}R^{ad}\right) 
\thickspace\textrm{.}
\end{eqnarray} 
The definition \refb{eqS8E} then leads to the classical entropy \cite{JacMye93}
\begin{equation}\labelb{eqI9}
S=\frac{1}{4}\int{(1+32\pi\alpha\,{}^{(n-2)}\! R)}
\thickspace\textrm{,}
\end{equation} 
where the integration can be made on any $(D-2)$-dimensional spacelike slice of the horizon and ${}^{(n-2)}\! R$ is curvature scalar computed with the induced metric on the slice.

The problem of microscopic description for this case was treated with Solodukhin's method by us \cite{Cvitan:2002cs}. This method allows to obtain a relation between  conformal charge and eigenvalue $\Delta$ but not their values independently. Consequently the relation between entropy derived with Cardy formula and classical entropy was a proportionality relation containing an unknown parameter. The method relied also essentially on particular assumptions like spherical symmetry.

Here we want to treat Gauss--Bonnet gravity with Carlip method, which is using Wald's covariant approach \cite{CovPhasSpac,Lee:nz,Iyer:1994ys} and is more suitable to generalizations.
We shall treat general black holes with Killing horizons without particular restrictions to spherical symmetry. We shall obtain separate values of conformal charge and eigenvalues of Hamiltonian. Also due to an interesting discussion about assumptions needed for these methods to be valid \cite{Soloviev98} 
and for these two methods to be consistent \cite{Carlip01} one is motivated to test the method for different interactions. Indeed in the present derivation for Gauss--Bonnet gravity we find consistency with Solodukhin method when the latter is amended in the sense of Carlip proposal \cite{Carlip01} as was done by us previously \cite{Cvitan:2002cs}.

\section{Covariant phase space methods and GB gravity}

The covariant phase space method was developed for a general diffeomorphism invariant field theory \cite{CovPhasSpac,Lee:nz,Iyer:1994ys}. For completeness we mention here the main steps. 
One starts with a Lagrangian $D$-form 
\begin{equation}
\mathbf{L}[\phi]=\bm{\epsilon}L[\phi]
\thickspace\textrm{,}
\end{equation} 
where $\phi$ is collection of fields, and $\bm{\epsilon}$ is volume $D$-form.
Variation of ${L}$ has the form
\begin{equation}
\delta\mathbf{L}[\phi]=\mathbf{E}[\phi]\delta\phi + d\mathbf{\Theta}\left[\phi,\delta\phi\right] 
\thickspace\textrm{.}
\end{equation} 
Here $\mathbf{E}$ is the equation of motion $D$-form and $\mathbf{\Theta}$ is $(D-1)$-form called symplectic potential. 
One introduces new vector field $\xi^a$ defining a diffeomorphism, and a corresponding  conserved current which is $(D-1)$-form given with
\begin{equation}
\mathbf{J}[\xi] = \mathbf{\Theta}[\phi,\mathcal{L}_\xi\phi] - \xi\cdot\mathbf{L}
\thickspace\textrm{,}
\end{equation} 
Dot denotes contraction. It was shown also \cite{Lee:nz} that 
\begin{equation}\labelb{eqJ8dQ}
\mathbf{J} = d\mathbf{Q}
\thickspace\textrm{,}
\end{equation} 
Hamiltonian equations of motion are then
\begin{equation}
\delta H[\xi]=\int_C
(\delta\mathbf{J}[\xi]-d(\xi\cdot\mathbf{\Theta}[\phi,\delta\phi]))
\thickspace\textrm{,}
\end{equation} 
where $C$ is a Cauchy surface.
Now using \refb{eqJ8dQ} one obtains
\begin{equation}\labelb{eqII7}
\delta H[\xi]=\int_{\partial C}
(\delta\mathbf{Q}[\xi]-\xi\cdot\mathbf{\Theta}[\phi,\delta\phi])
\thickspace\textrm{.}
\end{equation} 
Assuming\footnote{See \cite{Carlip99} for more details.} that a $(D-1)$-form $\mathbf{B}$, defined with
\begin{equation}\labelb{eqII8}
\delta\int_{\partial C}\xi\cdot\mathbf{B}=\int_{\partial C}\xi\cdot\mathbf{\Theta}
\thickspace\textrm{,}
\end{equation} 
exists, then \refb{eqII7} can be integrated to give
\begin{equation}\labelb{eqII9}
H[\xi]=\int_{\partial C}
(\mathbf{Q}[\xi]-\xi\cdot\mathbf{B})
\thickspace\textrm{.}
\end{equation} 
Thus, as stressed in Ref.~\cite{Iyer:1994ys}, Hamiltonian is a pure surface term when evaluated on shell. Here, diffeomorphism invariance was essential and not the particular form of equations of motion.

Consider two vector fields $\xi_1^a$, $\xi_2^a$. If fields $\phi$ solve the equations of motion, then one can show, due to vanishing of bulk constraints, that 
\begin{equation}\labelb{eqII10}
\delta_{\xi_2}H[\xi_1]=
\int_{\partial C} \left(\xi_2\cdot\mathbf{\Theta}[\phi,\mathcal{L}_{\xi_1}\phi]
   - \xi_1\cdot\mathbf{\Theta}[\phi,\mathcal{L}_{\xi_2}\phi] 
   - \xi_2\cdot(\xi_1\cdot{\bf L}) \right)
\thickspace\textrm{.}
\end{equation} 
Essential in forthcoming applications is to note that Poisson brackets of the generators $H[\xi]$ in the presence of boundaries can have central extension
\begin{equation}\labelb{eqII11}
\left\lbrace H[\xi_1], H[\xi_2] \right\rbrace =
H[\left\lbrace \xi_1, \xi_2 \right\rbrace] +
K[ \xi_1, \xi_2 ]
\thickspace\textrm{.}
\end{equation} 
Due to vanishing of bulk terms 
\begin{equation}\labelb{eqII12}
\delta_{\xi_2}H[\xi_1]=
\delta_{\xi_2}J[\xi_1]
\thickspace\textrm{,}
\end{equation} 
where $J$ is the boundary term. But Dirac bracket 
$\left\lbrace J[\xi_1], J[\xi_2] \right\rbrace^*$
is precisely $\delta_{\xi_2}J[\xi_1]$,
and as a consequence one has the algebra
\begin{equation}\labelb{eqII14}
{\left\lbrace J[\xi_1], J[\xi_2] \right\rbrace}^*
= J[\left\lbrace \xi_1, \xi_2 \right\rbrace] +
K[ \xi_1, \xi_2 ]
\thickspace\textrm{.}
\end{equation} 
Previous relations \refb{eqII10},\refb{eqII12} imply
\begin{equation}\labelb{eqII15}
{\left\lbrace J[\xi_1], J[\xi_2] \right\rbrace}^*=
\int_{\partial C} \left(
     \xi_2\cdot\mathbf{\Theta}[\phi,\mathcal{L}_{\xi_1}\phi]
   - \xi_1\cdot\mathbf{\Theta}[\phi,\mathcal{L}_{\xi_2}\phi] 
   - \xi_2\cdot(\xi_1\cdot{\bf L}) \right)
\thickspace\textrm{.}
\end{equation} 
In this paper, we are interested in Gauss--Bonnet Lagrangian which in particular contains no derivative of Riemann tensor. For such a case symplectic potential $\mathbf{\Theta}$ reads as follows \cite{Iyer:1994ys}
\begin{equation}
\Theta_{p_{\idxspc}\idxrange{a}{n-2}}=2\epsilon_{a_{\idxspc}p_{\idxspc}\idxrange{a}{n-2}}
(E_{R}^{abcd}\nabla_d\delta g_{bc}-\nabla_dE_{R}^{abcd}\delta g_{bc})
\thickspace\textrm{.}
\end{equation}
The tensor $E_{R}^{abcd}$ is already introduced in \refb{eqdefE}, and for particular Gauss--Bonnet Lagrangian it has the specific form \refb{eqdefEGB}. Due to Bianchi identity 
\begin{equation}
\nabla_{[e}R_{ab]cd} = 0
\thickspace\textrm{,}
\end{equation}
and antisymmetric properties of $\delta$ symbol in \refb{eqdefEGB}, one finds for Gauss--Bonnet case 
\begin{equation}\labelb{eqdivE}
\nabla_dE_{R}^{abcd} = 0
\thickspace\textrm{,}
\end{equation} 
Thus symplectic potential takes simple form
\begin{equation}
\Theta_{p_{\idxspc}\idxrange{a}{n-2}}=2\epsilon_{a_{\idxspc}p_{\idxspc}\idxrange{a}{n-2}}
E_{R}^{abcd}\nabla_d\delta g_{bc}
\thickspace\textrm{.}
\end{equation}
and 
\begin{eqnarray}\labelb{eqII23}
\left\lbrace J[\xi_1], J[\xi_2] \right\rbrace^*  = 
&2&\int_{\partial C}
\{
\epsilon_{a_{\idxspc}p_{\idxspc}\idxrange{a}{n-2}}
\left(
\xi_2^p E_{R}^{abcd}\nabla_d\delta_1 g_{bc}
-\xi_1^p E_{R}^{abcd}\nabla_d\delta_2 g_{bc}
\right) 
\nonumber \\
&&-\xi_2\cdot(\xi_1\cdot\bf{L})
\}
\thickspace\textrm{.}
\end{eqnarray} 

\section{Horizon as boundary}
In this section we shall impose existence of Killing horizon and consider a certain class of boundary conditions on it.
These conditions have been formulated by Carlip \cite{Carlip99} for Einstein gravity, and we shall assume that they hold also in the more general case treated in this paper. In particular we assume $D$-dimensional spacetime $M$ with boundary $\partial M$ such that we have a Killing vector $\chi^a$
\begin{equation}\labelb{eqIII1}
\chi^2 = g_{ab}\chi^a\chi^b = 0\thickspace\textrm{at $\partial M$}
\thickspace\textrm{.}
\end{equation} 
Near horizon ("stretched horizon") we define $\rho_a$
\begin{equation}\labelb{eqIII2}
\nabla_a\chi^2 = -2\kappa\rho_a
\thickspace\textrm{.}
\end{equation} 
It follows 
\begin{equation}
\chi^a\rho_a = 0
\thickspace\textrm{,}
\end{equation} 
\begin{equation}
\chi^a \rightarrow \rho^a \qquad 
\hbox{as $\chi^2\rightarrow 0$}
\thickspace\textrm{,}
\end{equation} 
but $\chi^a$ is timelike and $\rho^a$ is spacelike so that 
\begin{equation}\labelb{eqIII3}
\frac{\rho^2}{\chi^2}\rightarrow -1\qquad
\hbox{as $\chi^2\rightarrow 0$}
\thickspace\textrm{.}
\end{equation} 
We introduce variations that satisfy boundary conditions near the horizon as follows
\begin{equation}
{\chi^a\chi^b\over\chi^2}\delta g_{ab} \rightarrow 0 , \quad
\chi^a t^b\delta g_{ab} \rightarrow 0 \qquad 
\hbox{as $\chi^2\rightarrow 0$}
\thickspace\textrm{.}
\end{equation} 
In addition we require conditions on $\rho$ derivative
\begin{equation}
\rho^a\nabla_a(g_{bc}\delta g^{bc}) = 0 , \quad
\rho^a\nabla_a\left( {\rho^b\delta\chi_b\over\chi^2} \right) =
\rho^a\nabla_a\left( {\delta \rho^2\over\rho^2} \right) = 0 \quad
\hbox{at $\chi^2=0$}
\thickspace\textrm{,}
\end{equation} 
and we keep $\chi^a$ and $\rho_a$ fixed.
Here $t^a$ is any unit spacelike vector tangent to $\partial M$. We shall consider diffeomorphisms generated by vector fields $\xi^a$ where 
\begin{equation}\labelb{eqIII8}
\xi^a=T\chi^a+R\rho^a
\thickspace\textrm{,}
\end{equation} 
with conditions
\begin{equation}\labelb{eqIII9}
R=\frac{1}{\kappa}\frac{\chi^2}{\rho^2}\chi^a\nabla_a T\qquad 
\hbox{everywhere}
\thickspace\textrm{,}
\end{equation} 
and
\begin{equation}\labelb{eqIII10}
\rho^a\nabla_a T = 0\qquad 
\hbox{at horizon}
\thickspace\textrm{.}
\end{equation} 
It is easy to check that for a one parametric group of diffeomorphisms which satisfy \refb{eqIII8},\refb{eqIII9},\refb{eqIII10} the fields $\chi^a$ satisfy also
\begin{equation}
\{ \xi_1,\xi_2  \}^a = (T_1\dot{T}_2 - T_2\dot{T}_1)\chi^a +
  {1\over\kappa}{\chi^2\over\rho^2} {(T_1\dot{T}_2 - T_2\dot{T}_1)}\dot{} \rho^a
\thickspace\textrm{,}
\end{equation} 
which is a $\hbox{\it Diff\,}S^1$ or $\hbox{\it Diff\,}{\bf R}$ algebra ($\dot{T}\equiv\chi^a\nabla_aT$). 

An additional requirement will be necessary as already explained in \cite{Carlip99}
\begin{equation}\labelb{eqIII5.4}
\delta\int_{\partial C}{\bm{\hat\epsilon}}
   \left({\tilde\kappa} - {\rho\over|\chi|}\kappa\right) = 0 
\thickspace\textrm{,}
\end{equation} 
where ${\tilde\kappa}^2 = -a^2 / \chi^2$, and $a^a = \chi^b\nabla_b\chi^a$ is the acceleration of an orbit of $\chi^a$. This condition will guarantee existence of generators $H[\xi]$, and for diffeomorphisms \refb{eqIII8}, that satisfy 
$\dot{T}_\alpha=\lambda_\alpha T_\alpha$, will imply
\begin{equation}
\int_{\partial C} \bm{\hat\epsilon}\,
   T_\alpha T_\beta \sim \delta_{\alpha+\beta}
\thickspace\textrm{.}
\end{equation} 

Now we want to find the central term which from \refb{eqII14} can be calculated as
\begin{equation}
K[\xi_1,\xi_2] = \{ J[\xi_1], J[\xi_2] \}^* - J[\{\xi_1,\xi_2\}]
\thickspace\textrm{.}
\end{equation} 
In evaluating \refb{eqII15} we integrate over $(D-2)$-surface $\mathcal{H}$, which is the intersection of Killing horizon $\chi^2 = 0$ with the Cauchy surface $C$.
As usual we introduce two null normals on $\mathcal{H}$. One is Killing vector $\chi^a$ and the other is future directed null normal
$N^a = k^a -\alpha\chi^a - t^a$, 
where $t^a$ is tangent to $\mathcal{H}$ and has a norm $t^2=2\alpha-\alpha^2\chi^2$, and 
$k^a = -\left( \chi^a - \rho^a |\chi| / \rho \right) / \chi^2$.
Now the volume element can be written as 
\begin{equation}
\epsilon_{bca_1\dots a_{n-2}} = {\hat\epsilon}_{a_1\dots a_{n-2}}
\eta_{bc} + \dots
\thickspace\textrm{,}
\end{equation} 
where only the first term contributes to the integral, and binormal $\eta_{ab}$ is
\begin{equation}\labelb{eqeta}
\eta_{ab} = 2\chi_{[b}N_{c]}=\frac{2}{|\chi{}|\rho}\rho_{[a}\chi_{b]} + t_{[a}\chi_{b]} 
\thickspace\textrm{.}
\end{equation} 
Let us evaluate the first term in integral \refb{eqII15}.
Due to our boundary condition on horizon, the leading order term in $\chi^2$ is
\begin{eqnarray}
\nabla_d\delta g_{ab} 
& \equiv & \nabla_d\nabla_a\xi_b + \nabla_d\nabla_b\xi_a\\
& = & -2 \chi_d\chi_a\chi_b      {\ddot{T} \over \chi^4} + 
       2\chi_d\chi_{(a}\rho_{b)} 
             \left(   {\dddot{T} \over {\kappa\chi^2\rho^2}} +
                {{2 \kappa \dot{T}} \over \chi^4} \right) 
\thickspace\textrm{.}
\end{eqnarray} 
Also
\begin{eqnarray}
\xi_2^b\eta_{ab} & = & R_2\chi_a + T_2\rho_a + O(\chi^2)
\thickspace\textrm{.}
\end{eqnarray} 
Thus, the first term of \refb{eqII15} becomes
\begin{eqnarray}
E_R^{abcd}(R_2\chi_a + T_2\rho_a)
    \left( -2 \chi_d\chi_a\chi_b      {\ddot{T} \over \chi^4} + 
       2\chi_d\chi_{(a}\rho_{b)} 
             \left(   {\dddot{T} \over {\kappa\chi^2\rho^2}} +
                {{2 \kappa \dot{T}} \over \chi^4} \right) \right) 
\thickspace\textrm{.}
\end{eqnarray} 
At this stage it is useful to use symmetries of $E_R^{abcd}$ which are those of Riemann tensor so we have antisymmetry in $a\leftrightarrow b$ (and $c\leftrightarrow d$). For instance terms proportional to $\chi_a\chi_b\chi_c\chi_d$ will drop out and so many other.
We end with 
\begin{eqnarray}\labelb{eqIII23}
2E_R^{abcd}\chi_{[a}\rho_{b]}\chi_{[c}\rho_{d]}
\left(  {{T_2\dddot{T}_1} \over {\kappa\chi^2\rho^2}} +
  {{2 \kappa T_2 \dot{T_1}} \over \chi^4} \right) 
\thickspace\textrm{.}
\end{eqnarray} 
Using \refb{eqeta} and neglecting nonleading terms we get
\begin{eqnarray}
4E_R^{abcd}\chi_{[a}\rho_{b]}\chi_{[c}\rho_{d]} & = &
{|\chi{}|^2\rho^2}E_R^{abcd}\eta_{ab}\eta_{cd}
\thickspace\textrm{,}
\end{eqnarray} 
so that the first term of \refb{eqII15} becomes
\begin{equation}
\frac{1}{2}E_R^{abcd}\eta_{ab}\eta_{cd}
\left(  {2 \kappa T_2 \dot{T_1}} - {{T_2\dddot{T}_1} \over {\kappa}} \right)  + O(\chi^2)
\thickspace\textrm{.}
\end{equation}

The second term is exactly zero due to equation \refb{eqdivE} which is valid for Gauss--Bonnet gravity.

To evaluate contribution from third term in \refb{eqII23} we see using definition \refb{eqIII8} of vector field $\xi^a$ that
\begin{eqnarray}
\xi_2\cdot(\xi_1\cdot{\bf L}) & = &
{\chi^2 \over \kappa} \bm{\hat\epsilon} L (T_1 \dot{T}_2 - \dot{T}_1 T_2 ) = O(\chi^2)
\thickspace\textrm{,}
\end{eqnarray} 
since Lagrangian is finite on horizon.
Finally, we get
\begin{eqnarray}
\{ J[\xi_1], J[\xi_2] \}^* &=&
\frac{1}{2}
\int_{\mathcal{H}}
{\hat\epsilon}_{\idxrange{a}{n-2}}
E_R^{abcd}\eta_{ab}\eta_{cd} \times \nonumber \\ 
&{}&\times \left( {1 \over {\kappa}} ({T_1\dddot{T}_2}-{T_2\dddot{T}_1}) -
  {2 \kappa }(T_1 \dot{T_2} - T_2 \dot{T_1} \right)
\thickspace\textrm{.}
\end{eqnarray}

Next we need to calculate the Noether charge
\begin{eqnarray}\labelb{eqIII41}
Q_{\idxrangeb{c}{3}{n}}
&=&
-E_R^{abcd}\epsilon_{a_{\idxspc}b_{\idxspc}\idxrangeb{c}{3}{n}}
\nabla_{[c}\xi_{d]}
\thickspace\textrm{.}
\end{eqnarray}
Analogous procedure leads to
\begin{equation}
Q_{\idxrangeb{c}{3}{n}}
=
-{1 \over 2}E_R^{abcd}\eta_{ab}\eta_{cd}
\left( 2 \kappa T - {\ddot{T} \over {\kappa}} \right)
\hat\epsilon_{\idxrangeb{c}{3}{n}}
\thickspace\textrm{.}
\end{equation}
Using the same method, we can calculate
\begin{eqnarray}
J[\{\xi_1,\xi_2\}] &=&
-{1 \over 2}\int_{\mathcal{H}}{\hat\epsilon}_{\idxrange{a}{n-2}}
E_R^{abcd}\eta_{ab}\eta_{cd} \nonumber \\
&{}&
\left( {2 \kappa }(T_1 \dot{T_2} - T_2 \dot{T_1}) - 
  {1 \over {\kappa}} (\dot{T}_1\ddot{T}_2-\ddot{T}_1\dot{T}_2+
                      T_1\dddot{T}_2-\dddot{T}_1 T_2)
  \right)
\thickspace\textrm{.}
\end{eqnarray}
Now we are able to deduce central charge from \refb{eqII14}
\begin{eqnarray}\labelb{eqIII44}
K[\xi_1,\xi_2] &=&
-{1 \over 2}\int_{\mathcal{H}}{\hat\epsilon}_{\idxrange{a}{n-2}}
E_R^{abcd}\eta_{ab}\eta_{cd}
  {1 \over {\kappa}} (\dot{T}_1\ddot{T}_2-\ddot{T}_1\dot{T}_2)
\thickspace\textrm{.}
\end{eqnarray}

\section{Conformal charge and entropy}
In previous sections we have introduced constraint algebra \refb{eqII14} where we have calculated various terms.
As explained in \cite{Carlip99}, this algebra can be connected to the Virasoro algebra of diffeomorphisms of the circle or the real line.
One introduces another condition which will ensure that \refb{eqII14} indeed coincides with Virasoro algebra. For this purpose we define the parameter $v$ of the orbits of the Killing vector $\chi^a\nabla_a v=1$ and consider functions $T_1$,$T_2$ of $v$ and angular coordinates on horizon such that they satisfy
\begin{eqnarray}\labelb{eqIV1}
{1 \over A}
\int_{\mathcal{H}}\bm{\hat\epsilon}\,
T_1(v,\theta)T_2(v,\theta)
&=&
{\kappa' \over {2\pi}}
\int d v 
T_1(v,\theta)T_2(v,\theta)
\thickspace\textrm{,}
\end{eqnarray}
where $A = \int_{\mathcal{H}}\bm{\hat\epsilon}$ is the area of the horizon and $2\pi/\kappa'$ is period in the variable $v$ of the functions $T(v,\theta)$.
In fact we can be more specific. For rotating black hole
\begin{equation}
\chi^a = t^a + \sum{\Omega_{i} \psi_i^a}
\thickspace\textrm{,}
\end{equation} 
where $t^a$ is time translation Killing vector, $\psi_i^a$ are rotational Killing vectors with corresponding angles $\psi_i$ and angular velocities $\Omega_{i}$. We shall sometimes, instead of variables $t$, $\psi_i$ connected with orbits of 
$t^a$, $\psi_i^a$, work with variables $(v,\theta_i)$ connected with 
orbits of $\chi^a$, $\theta_i^a = \psi_i^a$. Then $v=t$, $\theta_i = \psi_i - \Omega_i v$, and we choose for diffeomorphism defining functions $T_n$
\begin{eqnarray}
T_n&=&{1 \over {\kappa'}}e^{in\left(\kappa'v+\sum\phi_i(\psi_i-\Omega_i v)\right)} \\
   &=&{1 \over {\kappa'}}e^{in\left(\kappa'v+\sum\phi_i\theta_i\right)}
\thickspace\textrm{.}
\end{eqnarray} 
These functions are of the form
\begin{eqnarray}\labelb{eqIV4}
T_n(v,\theta_i)&=&{1 \over {\kappa'}}e^{in\kappa'v}f_n(\theta_i)
\thickspace\textrm{.}
\end{eqnarray} 
They satisfy
\begin{equation}
{1 \over A}\int\bm{\hat\epsilon}\, T_n T_m = \delta_{n+m,0}{1 \over {{\kappa'}^2}}
\thickspace\textrm{,}
\end{equation}
and in particular
\begin{equation}
{1 \over A}\int\bm{\hat\epsilon} f_n f_m = \delta_{n+m,0}
\thickspace\textrm{.}
\end{equation}
At this point classical Virasoro condition can be checked in the form
\begin{equation}
\{T_m,T_n\} = -i (m-n)T_{m+n}
\thickspace\textrm{.}
\end{equation}
Also we see that condition \refb{eqIV1} is fullfilled and thus enables us to obtain full Virasoro algebra with nontrivial central term $K[T_m,T_n]$ which can be calculated from \refb{eqIII44}
\begin{equation}
iK[T_m,T_n] = \left({\kappa' \over \kappa}\right){\hat{A} \over 8\pi} m^3 \delta_{m+n,0}
\thickspace\textrm{,}
\end{equation}
where
\begin{equation}
\hat{A} \equiv -8\pi
\int_{\mathcal{H}}{\hat\epsilon}_{\idxrange{a}{n-2}}
E_R^{abcd}\eta_{ab}\eta_{cd}
\thickspace\textrm{.}
\end{equation}
Here, we have used that metric does not depend on variables $\mathbf{\theta}_i$ on which diffeomorphism defining functions $T_n$ depend. That enabled us to factorize the integral in \refb{eqIII44}.
Finally, we obtain Virasoro algebra
\begin{equation}
i \{ J[\xi_1], J[\xi_2] \}^*
= (m-n)J[T_{m+n}]+\frac{c}{12}m^3 \delta_{m+n,0}
\thickspace\textrm{,}
\end{equation}
and central charge $c$ is equal to
\begin{equation}
\frac{c}{12}=\frac{\hat{A}}{8\pi}\frac{\kappa'}{\kappa}
\thickspace\textrm{.}
\end{equation}
Now, we want to calculate the value of the Hamiltonian. This is given with the first term in relation \refb{eqII9} where the second term can be neglected\footnote{
As in Einstein case, condition \refb{eqIII5.4} enables us to factorize $\xi\cdot\mathbf{\Theta}$ into
$\frac{1}{2}E_R^{abcd}\eta_{ab}\eta_{cd}\times\delta(\textrm{terms that vanish on shell})$, which together with \refb{eqII8}
implies that $\int_{\mathcal{H}}\xi\cdot\mathbf{B}$ vanishes on shell.}.
This first term can be calculated from \refb{eqIII41} and 
\begin{equation}
T_0 = \frac{1}{\kappa'}
\thickspace\textrm{.}
\end{equation}
Thus, 
\begin{equation}
J[T_0] = - \int_{\mathcal{H}}{\hat\epsilon}_{\idxrange{a}{n-2}}
E_R^{abcd}\eta_{ab}\eta_{cd}
\frac{\kappa}{\kappa'}
\thickspace\textrm{,}
\end{equation}
or
\begin{equation}
\Delta \equiv J[T_0] = \frac{\kappa}{\kappa'}\hat{A}
\thickspace\textrm{.}
\end{equation}

Before analyzing Cardy formula we note following relations between conformal charge and eigenvalue $\Delta$
\begin{equation}
\frac{c \Delta}{12} = \left(\frac{\hat{A}}{8\pi}\right)^2
\thickspace\textrm{,}
\end{equation}
or 
\begin{equation}
\frac{1}{12}\frac{c}{\Delta} = \left(\frac{\kappa'}{\kappa}\right)^2
\thickspace\textrm{.}
\end{equation}
We are interested in calculating entropy via Cardy formula \refb{eqI1}. Thus
\begin{equation}
S = \frac{\hat{A}}{4}\sqrt{2-\left(\frac{\kappa'}{\kappa}\right)^2}
\thickspace\textrm{.}
\end{equation}
Thus entropy is proportional to the classical entropy \refb{eqS8E}. The constant of proportionality is dimensionless. The proportionality relation becomes equality when we take for the period of functions $T_n$ the period of the Euclidean black hole \cite{Carlip99}, \cite{EuclPeriod}, \cite{Cvitan:2002cs}.

In that case 
\begin{equation}\labelb{eqIV25}
S = \frac{\hat{A}}{4}=-2\pi\int{E_{R}^{abcd}\eta_{ab}\eta_{cd}}
\thickspace\textrm{,}
\end{equation}
and also $c/12=\Delta$.
This formula can be written in more explicit form using specific properties of Gauss--Bonnet gravity. In fact for Gauss--Bonnet gravity it was shown that 
\begin{equation}
E_{R}^{abcd}\eta_{ab}\eta_{cd} = -\frac{1}{8\pi}-4\alpha\,{}^{(n-2)}\! R
\thickspace\textrm{,}
\end{equation}
where ${}^{(n-2)}\! R$ is curvature scalar computed with the induced metric on horizon.
This relation was demonstrated using Gauss--Codazzi relation
\begin{equation}\labelb{eqGC}
{}^{(n-2)}\! R = {}^{(n)}\! R - 2t^{ab}R_{ab}+t^{ac}t^{bd}R_{abcd} + \dots
\thickspace\textrm{,}
\end{equation}
where dots denote "external curvature terms".
These additional terms can be neglected due to the fact that bifurcation surface can be chosen for integration and the fact that the integral is the same for any section of a stationary horizon \cite{genent}.
Here,
\begin{equation}
t_{ab} = -n_a n_b + r_a r_b
\thickspace\textrm{}
\end{equation}
is the metric for $\mathcal{H}$ and $n^a$, $r^a$ are unit timelike and spacelike normals.
Now,
\begin{eqnarray}
E_{R}^{abcd}\eta_{ab}\eta_{cd}
 &=& -\frac{1}{8\pi}-4\alpha(R + 2R^{ac}\eta_{ad}\eta_{cd}-\frac{1}{2}R^{abcd}\eta_{ab}\eta_{cd})\\
 &=& -\frac{1}{8\pi}-4\alpha\,{}^{(n-2)}\! R
\thickspace\textrm{.}
\end{eqnarray}
Here, first equality is due to explicit form of Gauss--Bonnet Lagrangian and second equality is due to identity $\eta_{ad}\eta_{cd}g^{bd} = -t_{ac}$ and Gauss--Codazzi relation \refb{eqGC}.
Thus,
\begin{equation}
\hat{A} = A + 32\pi \alpha \int_{\mathcal{H}}\bm{\hat\epsilon}\;{}^{(n-2)}\! R
\thickspace\textrm{,}
\end{equation}
and expression \refb{eqIV25} for entropy becomes
\begin{equation}
S = \frac{A}{4} + 8\pi \alpha \int_{\mathcal{H}}\bm{\hat\epsilon}\;{}^{(n-2)}\!R
\thickspace\textrm{,}
\end{equation}
This is exactly the classical result \refb{eqI9}. Similarly for general GB case the expression \refb{eqIV25} can also be written as \cite{Visser:1993nu}.
\begin{equation}
S = -4\pi \sum_{m=1}^{[\frac{D}{2}]}m \lambda_m \int \bm{\hat\epsilon}\; L_{m-1}
\thickspace\textrm{,}
\end{equation}
where $L_m$ is evaluated using induced metric on $\mathcal{H}$.

\section{Conclusion}
In this paper we have tried to make a progress in the efforts to give microscopic interpretation to entropy formulas for more general theories then Einstein theory. Here, the $D$-dimensional extended Gauss--Bonnet theory was considered. It was shown that using Carlip method \cite{Carlip99} and asking certain boundary conditions near black hole horizon one can define an algebra of diffeomorphisms containing Virasoro algebra as its subalgebra. 

Calculation of central charge enables, with the help of Cardy formula, to find the entropy which is as expected different than area law but agrees with Gauss--Bonnet entropy as derived in \cite{Iyer:1994ys}, \cite{JacMye93}.

The result is more general then alternative derivation by us in reference \cite{Cvitan:2002cs}, because here we do not have to restrict ourselves to spherical symmetry. Also here it is possible to calculate separately central charge and eigenvalue of Virasoro generator $L_0$. It is also encouraging that it shows that two methods give consistent results.

\section*{Acknowledgements}
We would like to acknowledge the financial support
under the contract No.~0119261 of Ministry of Science and Technology of Republic of
Croatia.


\begin{thebibliography}{99}

\bibitem{Strominger:1996sh}
A.~Strominger and C.~Vafa, Phys.\ Lett.\ B {\bf 379} (1996) 99, hep-th/9601029;\\
J.~R.~David, G.~Mandal and S.~R.~Wadia, Phys.\ Rept.\  {\bf 369} (2002) 549, hep-th/0203048;\\
D.~Amati, J.~G.~Russo, Phys.\ Lett.\ B {\bf 454} (1999) 207, hep-th/9901092.

\bibitem{Ashtekar:1997yu}
A.~Ashtekar, J.~Baez, A.~Corichi and K.~Krasnov, Phys.\ Rev.\ Lett.\  {\bf 80} (1998) 904, gr-qc/9710007;\\
A.~Ashtekar, C.~Beetle and S.~Fairhurst, Class.\ Quant.\ Grav.\  {\bf 16} (1999) L1, gr-qc/9812065;\\
A.~Ashtekar, J.~C.~Baez and K.~Krasnov, Adv.\ Theor.\ Math.\ Phys.\  {\bf 4} (2000) 1, gr-qc/0005126;\\
A.~Ashtekar, A.~Corichi and K.~Krasnov, Adv.\ Theor.\ Math.\ Phys.\  {\bf 3} (2000) 419, gr-qc/9905089.

\bibitem{BroHen86}
J. D. Brown and M. Henneaux, Commun.\ Math.\ Phys.\  {\bf 104} (1986) 207.


\bibitem{BanStro}
M. Banados, C. Teitelboim and J. Zanelli, Phys.\ Rev.\ Lett.\  {\bf 69} (1992) 1849, hep-th/9204099;\\
A. Strominger, JHEP {\bf 9802} (1998) 009.

\bibitem{Solod99}
S. N. Solodukhin, Phys.\ Lett.\ B {\bf 454} (1999) 213, hep-th/9812056;

\bibitem{Carlip02}
S. Carlip, Phys.\ Rev.\ Lett.\  {\bf 88} (2002) 241301,  gr-qc/0203001.

\bibitem{Carlip01}
S.~Carlip, Phys.\ Lett.\ B {\bf 508} (2001) 168, gr-qc/0103100.


\bibitem{Carlip98}
S.~Carlip, Phys.\ Rev.\ Lett.\  {\bf 82} (1999) 2828, hep-th/9812013.

\bibitem{Carlip99}
S.~Carlip, Class.\ Quant.\ Grav.\  {\bf 16} (1999) 3327, gr-qc/9906126.

\bibitem{Cardy}
J. A. Cardy, Nucl.\ Phys.\ B {\bf 270} (1986) 186;\\
H. W. J. Bl\"{o}te, J. A. Cardy and M. P. Nightingale, Phys.\ Rev.\ Lett.\  {\bf 56} (1986) 742.

\bibitem{BekHaw}
J. D. Bekenstein, Lett.\ Nuovo. Cim.\  {\bf 4} (1972) 737;\\
J. D. Bekenstein, Phys.\ Rev. D\  {\bf 7} (1973) 2333;\\
J. D. Bekenstein, Phys.\ Rev. D\  {\bf 9} (1974) 3292;\\
S. W. Hawking, Nature {\bf 248} (1974) 30;\\
S. W. Hawking, Commun.\ Math.\ Phys.\  {\bf 43} (1975) 199.


\bibitem{Silva02}
S.~Silva, Class.\ Quant.\ Grav.\  {\bf 19} (2002) 3947, hep-th/0204179;\\
J.~l.~Jing and M.~L.~Yan, Phys.\ Rev.\ D {\bf 62} (2000) 104013, gr-qc/0004061;\\
H.~Terashima, Phys.\ Rev.\ D {\bf 64} (2001) 064016, hep-th/0102097.

\bibitem{Gour:2002pj}
G.~Gour, gr-qc/0210024;\\
J.~D.~Bekenstein and G.~Gour, Phys.\ Rev.\ D {\bf 66} (2002) 024005, gr-qc/0202034.

\bibitem{Jing:2000yn}
J.~l.~Jing and M.~L.~Yan, Phys.\ Rev.\ D {\bf 63} (2001) 024003, gr-qc/0005105



\bibitem{Lovelock}
D. Lovelock, J.\ Math.\ Phys.\  {\bf 12} (1971) 498;\\
D. Lovelock, J.\ Math.\ Phys.\  {\bf 13} (1972) 874;\\
R. C. Myers and J. Z. Simon, Phys.\ Rev.\ D\  {\bf 38} (1988) 2434;\\
R. G. Cai, Phys.\ Rev.\ D\  {\bf 65} (2002) 084014, hep-th/0109133;\\
R. G. Cai and K. S. Soh, Phys.\ Rev.\ D\  {\bf 59} (1999) 044013, gr-qc/9808067;\\
M. Cveti\v{c}, S. Nojiri and S. D. Odintsov, Nucl.\ Phys.\ B {\bf 628} (2002) 295, hep-th/0112045.

\bibitem{genent}
R. M. Wald, Phys.\ Rev.\ D {\bf 48} (1993) 3427, gr-qc/9307038;\\
T. Jacobson, G. Kang and R. C. Myers, Phys.\ Rev.\ D {\bf 49} (1994) 6587, gr-qc/9312023.

\bibitem{Iyer:1994ys}
V.~Iyer and R.~M.~Wald, Phys.\ Rev.\ D {\bf 50} (1994) 846, gr-qc/9403028.

\bibitem{JacMye93}
T.~Jacobson and R.~C.~Myers, Phys.\ Rev.\ Lett.\  {\bf 70} (1993) 3684, hep-th/9305016.

\bibitem{Cvitan:2002cs}
M.~Cvitan, S.~Pallua and P.~Prester, Phys.\ Lett.\ B {\bf 546} (2002) 119, hep-th/0207265.



\bibitem{CovPhasSpac}
C.~Crnkovi\'{c} and E.~Witten,  in: S.~W.~Hawking and W.~Israel, eds., {\it Three hundred years of gravitation} (Cambridge Univ. Press, 1989) 676-684;\\
C.~Crnkovi\'{c}, Class.\ Quant.\ Grav.\  {\bf 5} (1988) 1557;\\
E.~Witten, Nucl.\ Phys.\ B {\bf 276} (1986) 291;\\
G.~J.~Zuckerman, in: S.~T.~Yau, ed., {\it Mathematical aspects of string theory, Proceedings, San Diego 1986}, Adv.\ Ser.\ Math.\ Phys.\  {\bf 1} (1987) 259-284;\\
B.~Julia and S.~Silva, hep-th/0205072.



\bibitem{Lee:nz}
J.~Lee and R.~M.~Wald, J.\ Math.\ Phys.\  {\bf 31}, 725 (1990).


\bibitem{Soloviev98}
V. O. Soloviev, Phys.\ Rev.\ D {\bf 61} (1999) 027502, hep-th/9905220.\\
M.-I. Park, Nucl.\ Phys.\ B {\bf 634} (2002) 339, hep-th/0111224;\\
S.~Carlip, Phys.\ Rev.\ Lett.\  {\bf 83} (1999) 5596, hep-th/9910247.



\bibitem{EuclPeriod}
A.~Barvinsky, S.~Das and G.~Kunstatter, hep-th/0209039;\\
H.~A.~Kastrup, Phys.\ Lett.\ B {\bf 385} (1996) 75, gr-qc/9605038;\\
J.~Louko and J.~Makela, Phys.\ Rev.\ D {\bf 54} (1996) 4982, gr-qc/9605058;\\
J.~Makela and P.~Repo, Phys.\ Rev.\ D {\bf 57} (1998) 4899, gr-qc/9708029;\\
M.~Bojowald, H.~A.~Kastrup, F.~Schramm and T.~Strobl, Phys.\ Rev.\ D {\bf 62} (2000) 044026, gr-qc/9906105.

\bibitem{Visser:1993nu}
M.~Visser, Phys.\ Rev.\ D {\bf 48} (1993) 5697, hep-th/9307194.

\end{thebibliography}
\end{document}